# Optimization of the sensitivity of a temperature sensor based on the germanium-vacancy color center (GeV) in diamond

I.S. Cojocaru[1,2,3], V.V. Soshenko[2,3], S.V. Bolshedvorskii[2,3], V.A. Davydov[4], L. F. Kulikova[4], V. N. Agafonov[5], A. Chernyavskiy[2,6], A.N. Smolyaninov[3], V.N. Sorokin[1,2], S.Ya. Kilin[7,8], A.V. Akimov[1,2,3]

[1]*Russian Quantum Center, Bolshoy Boulevard 30, building 1, Moscow, 143025, Russia*

[2]*P.N. Lebedev Institute RAS, Leninsky Prospekt 53, Moscow, 119991, Russia*

[3]*LLC Diamond Sensors, 125130 Narvskaya St. 1A, Moscow, Russia*

[4]*Vereshchagin Institute for High Pressure Physics, Russian Academy of Sciences, Troitsk, Moscow, 108840, Russia*

[5]*GREMAN, CNRS, UMR 7347, INSA CVL, Université de Tours, Tours 37200, France*

[6]*Moscow Institute of Physics and Technology, 9 Institutskiy per., Dolgoprudny, Moscow Region, 141701, Russia*

[7]*National Research Nuclear University "MEPhI , 31, Kashirskoe Highway, Moscow, 115409 Russia*

[8]*B.I. Stepanov Institute of Physics NASB, 68, Nezavisimosty Ave, Minsk, 220072 Belarus*

*email: a.akimov@rqc.ru*

Temperature sensors based on the GeV color center in diamond are gaining considerable attention in both scientific and industrial fields. For widespread industrial adoption, however, these sensors need a design that is as simple and cost-effective as possible. The original sensor design relied on measuring the spectral characteristics of the zero-phonon line. Recently, a modified approach was introduced, which involves splitting the GeV emission with a dichroic mirror and determining temperature based on the ratio of the two resulting signals. In this analysis, we provide a detailed comparison of both methods. At room temperature, the two methods show comparable performance, with slight variations depending on component quality. However, at temperatures around 300 °C, the new method's performance is estimated to be nearly twice that of the original, provided optimal filter parameters are used. Additionally, the sensitivity of the new method remains roughly consistent with its performance at room temperature.

## I. INTRODUCTION

Temperature measurements at the nano- and microscale are crucial in numerous fields, including quantum physics, microelectronics, and biomedicine [1,2]. Among diamond-based temperature sensors, GeV color centers, in particular, are attracting significant attention due to their relatively high sensitivity and accuracy, combined with high spatial resolution [3]. Additionally, the chemical and physical inertness of diamond makes this sensor suitable for minimally invasive measurements and for use in harsh environments.

The initial method of temperature measurement with GeV centers employed a spectrometer to track either the position or width of the zero-phonon line (ZPL) as a temperature indicator [3–6]. This approach enabled temperature measurements within a valuable range of (-100 °C;600 °C) [7] and could even be realized as a fiber-based sensor [6]. This is beneficial in scenarios where electronic devices cannot be positioned directly at the measurement site, such as in biological applications or extreme environments. Various methods for the as-grown production of GeV centers in diamond have been proposed, making reasonably affordable sensing elements more accessible [7–13]. However, these measurement methods require complex equipment, like a spectrometer, and involve intricate spectral analysis, which can make scaling up the sensor challenging. A more recent approach [14] suggested replacing the spectrometer with a dichroic mirror, simplifying the sensor's construction and substituting complex mathematical fitting procedures with straightforward division. This method, however, requires precise calibration of the filter's central wavelength and employs costly sharp transmission/reflection slope filters for the thermometer. Additionally, the choice of central wavelength depends on the target temperature range.

In this work, we further develop this concept by examining the effects of the filter slope, its central wavelength, and transmission/reflection windows on the device's sensitivity across a practically important temperature range: 25 °C to 300 °C. This range includes biomedical and industrial applications, such as using optical thermometers to measure temperatures in high-voltage oil transformers (which can reach up to 300 °C). We demonstrate that it is possible to achieve equivalent sensitivity with a more compact and straightforward device. This approach has the potential to reduce the cost of thermometers based on GeV centers in diamond by making them more compact. Compared to spectrometer-based devices, it can also provide faster operation with the same level of sensitivity. Furthermore, additional sensitivity improvements can potentially be achieved by applying traditional noise suppression techniques, such as lock-in detection with a current-modulated laser diode.

## II. TRADITIONAL GEV THERMOMETRY

As a starting point for the research, we measured the traditional GeV spectrum using a custom-built confocal setup with the capability to control diamond temperature, as shown in Figure 1A. The confocal setup has position feedback based on lock-in detector (frequencies are 119, 186, 17 Hz for x, y and z correspondently), operating only when laser spot is fixed at one point using Cambridge Technology 6215H Galvanometer Scanner together with Physik Instrumente P-721.CDQ scanner and PerkinElmer avalanche photodiode. For spectral analysis, we used a SOLAR LS M266 spectrometer equipped with a Tucsen Dhyana 400D camera. The GeV center was excited by a Compass 315M-100 laser, with approximately 1.1 mW of laser power reaching the diamond through an Olympus Plan N 10X objective lens with a numerical aperture (NA) of 0.22.

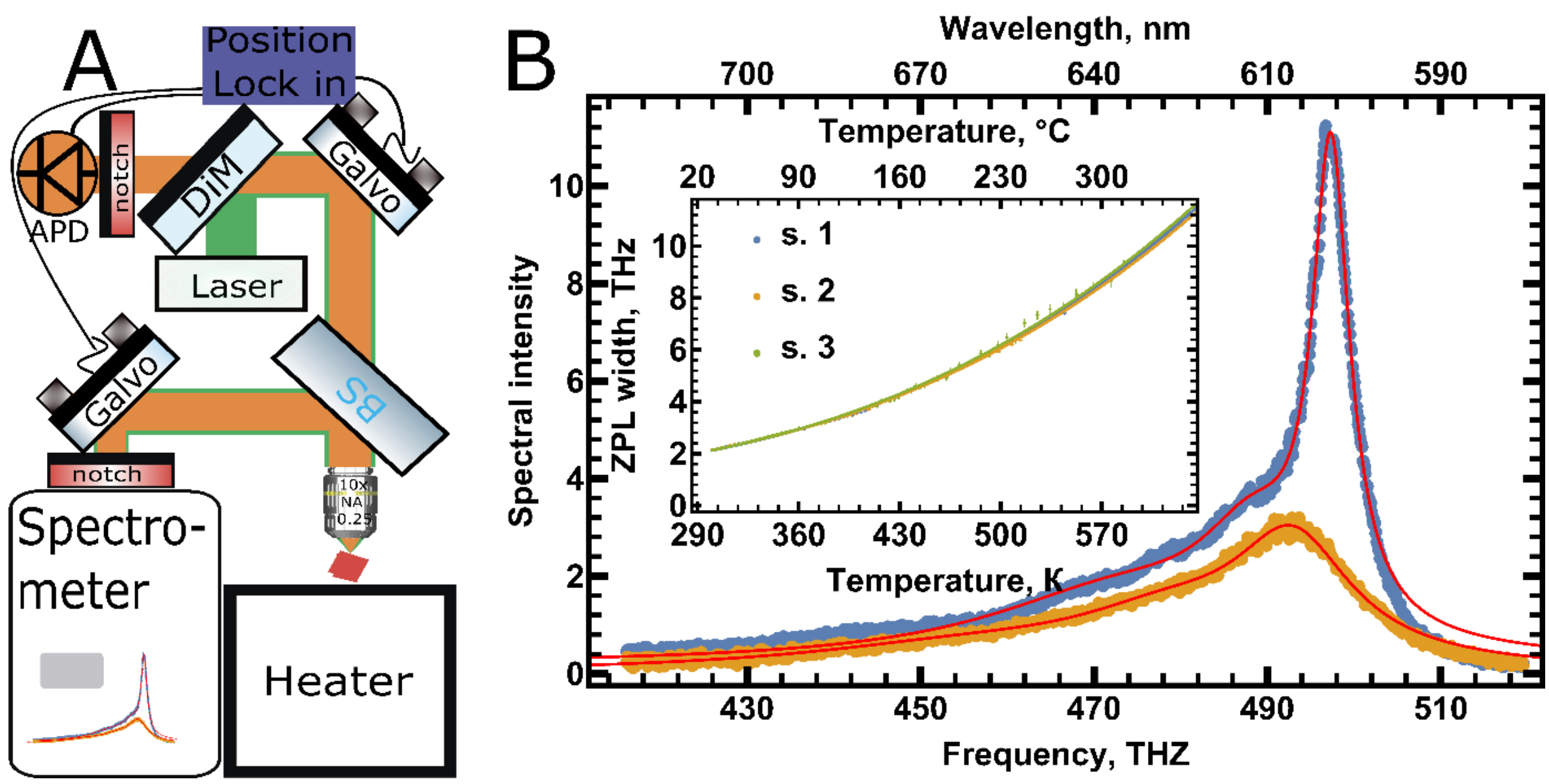

*Figure 1 A) Schematic of the experimental setup for spectrometer-based temperature measurements. APD – avalanche photodiode, DiM – dichroic mirror, BS – beam splitter, notch – notch filter, Lock in – stabilization of the beam position using a digital lock-in amplifier feedback loop on a specific position on micro-diamond using galvo mirrors (Galvo), Laser – CW green laser (532 nm). B) Example GeV spectra at 55 °C*

*(blue) and 276 °C (orange) fitted with a sum of three Lorentzian functions plus background (red line). The inset shows the temperature dependence of the ZPL width for three different microdiamonds (s. 1-3).*

A ceramic heater was used to control the temperature, equipped with two thermocouples connected to a MAX31855-K converter. The heater's temperature was stabilized using a digital feedback loop with NI 6733 and NI 6602 boards and a custom proportional-integral controller. To prevent uncontrolled temperature gradients, a Testo 865 thermovision camera monitored the heater's temperature, and the heater was mounted vertically to minimize convection effects. A 500 μm diameter and 1.5 mm deep hole was drilled into the heater to hold the microdiamonds with GeV centers. The laser beam, after passing through the objective lens, propagated through multiple layers of aluminum foil, creating an isolated air region around the microdiamond.

Micrometer-sized diamonds with GeV centers were synthesized using the high-pressure high-temperature (HPHT) method, following a catalyst metal-free halogenated hydrocarbon growth system as described in [3]. In this setup, GeV spectra were measured for temperatures ranging from 25 to 370 °C for three microdiamonds (~10 μm in size, samples 1–3 in Figure 1B), with an integration time of 7 seconds per spectrum. The spectra were fitted using the minimum mean square error method, with a model consisting of a sum of three Lorentzian functions and a background (two functions for the phonon sideband and one for the ZPL). Changes in the half-width at half-maximum (HWHM) were extrapolated using a power law, following the approaches in [4] and [7]. The fitting parameters closely aligned with previously published data (Figure 1B). Additionally, to assess sensitivity, 300 spectra were recorded at temperatures of 55 °C and 276 °C.

It is important to note that our setup was not optimized to achieve record sensitivity levels due to the division into two channels, low laser intensity, and the objective's numerical aperture. Consequently, temperature measurements based on fitting the ZPL HWHM yielded a sensitivity of 0.8 °C at T=55 °C and 2.9 °C at T=276 °C, with a spectrum integration time of 7 s (Figure 6). This sensitivity was evidently limited by the low signal, as confirmed by increasing the laser power (with the addition of a pulsed green laser), which resulted in sensitivity following a square-root dependence on the total pump intensity. Nevertheless, [14] has demonstrated that it is possible to experimentally achieve $\sim 30\ \mathrm{mK}/\sqrt{\mathrm{Hz}}$ sensitivity at room temperature using a dichroic filter and a spectral range cutoff filter. Thus, it would be valuable to examine the sensitivity dependence on the parameters of the measurement scheme presented below.

## III. SIMPLIFIED MEASUREMENT SCHEME AND OPTIMAL FILTER SELECTION.

In this work, we explore simplifying the measurement scheme by replacing the spectrometer with a dichroic mirror and two photodiodes, as illustrated in Figure 2. A green laser diode operating in constant intensity mode with a fiber output connected via a fiber optical circulator is proposed as the pump source. It's worth noting that standard methods for combining laser radiation and fluorescence, such as a dichroic mirror or beam splitter (including fiber- or polarization-based), are also applicable. The key innovation here is the use of a dichroic mirror that operates on the slope of the GeV ZPL, enabling temperature determination based on the intensity ratio at the arms $R(T)$ between parts of the GeV spectra.

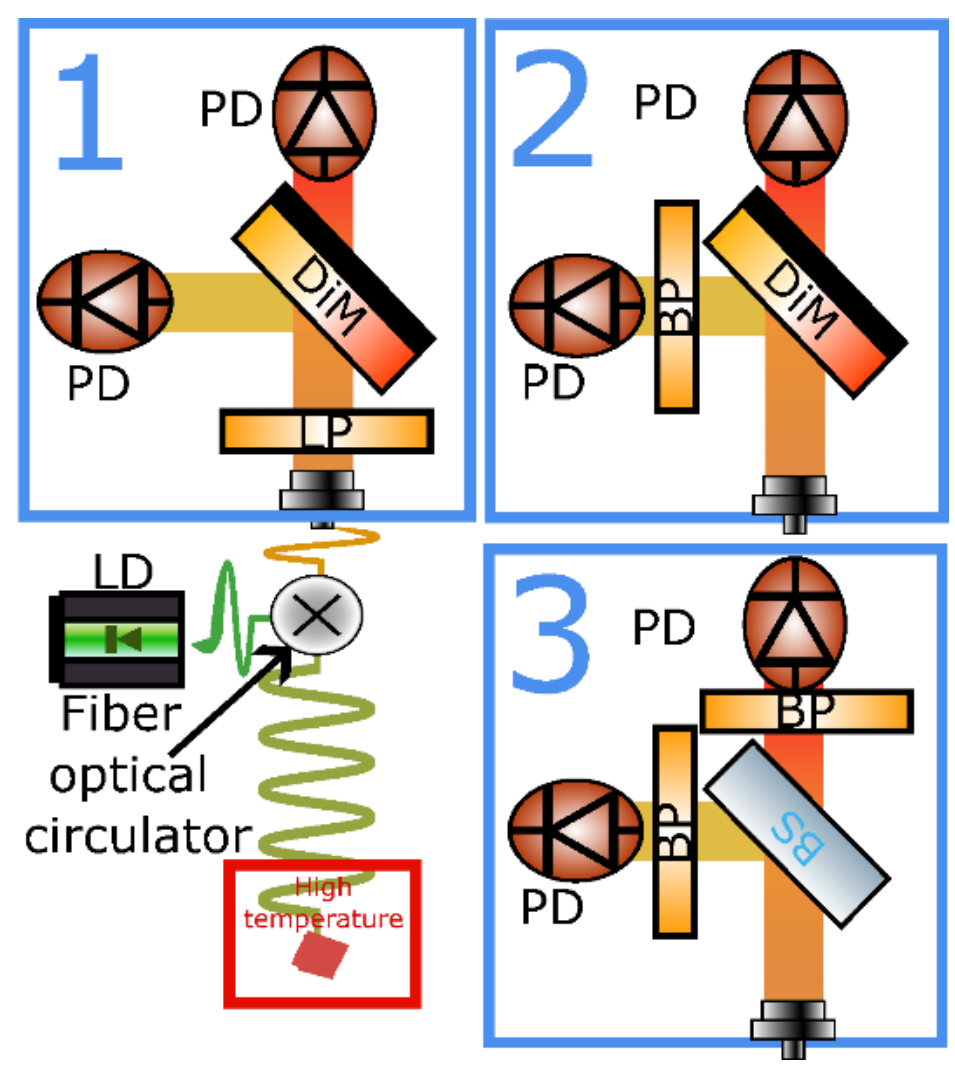

*Figure 2 Simplified diagram of a temperature sensor based on GeV color centers in diamond. LD – laser diode; LP – long-pass filters; BP – band-pass filter; DiM – dichroic mirror; BS – beam splitter, PD – photodiode. Three measurement configurations were considered in this work: 1) a dichroic mirror with a long-pass filter, 2) a dichroic mirror with a bandpass filter, and 3) two bandpass filters.*

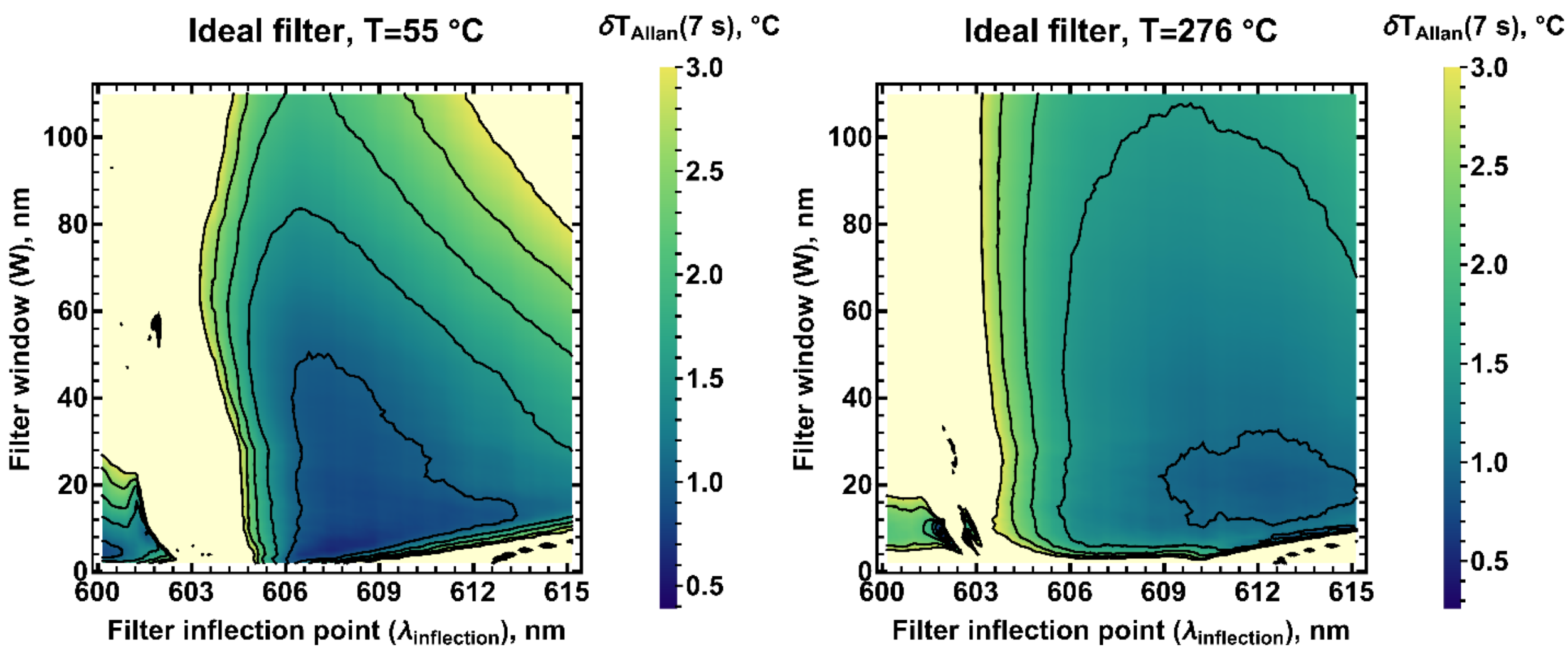

*Figure 3 Allan deviation at 7 s spectra integration time ( $\partial T_{Allan}(7\text{s})$ ) as a function of the parameters of the ideal filter for sample No. 2. The contour is plotted for every 0.5 °C.*

To validate this method, we used the same spectral data as described earlier. To minimize laser noise interference, it is advisable to use a filter (long-pass filter in Figure 2-1) that blocks the laser while incorporating scattering from both the diamond and the fiber. Consequently, data from spectra with wavelengths greater than 577 nm were analyzed to optimize the parameters of the dichroic mirror. Initially, we considered an ideal filter that measures the total signal within a specified window (W) on both the left and right sides of the spectrum's split point, corresponding to the inflection point of filter transmission. $\lambda_{\text{inflection}}$ :

$$R_{ideal}(T) = \sum_{\lambda_{\text{inflection}} \le \lambda \le \lambda_{\text{inflection}} + W} I(\lambda) \Bigg/ \sum_{\lambda_{\text{inflection}} - W \le \lambda < \lambda_{\text{inflection}}} I(\lambda) \qquad (1)$$

The resulting dependence was approximated by a third-degree polynomial $p(3, R(T)): R(T) \to T$ and the mean measurement error was calculated for it: $\Delta T_{mean} = \overline{\left| p(3, R(T)) - T \right|}$. To compare the sensitivity of various measurement methods, we employed Allan deviation, a metric routinely used to assess the stability of optical clocks [15–17], as well as diamond-based magnetic [18] and temperature sensors [19]. Typically, the maximum possible sensitivity is taken at the minimum of an Allan deviation, while value at 1 s defines sensitivity per 1 Hz bandwidth. For relative comparison here, we used the Allan deviation at 7 s, corresponding to the integration time used in our experiment, as a measure of "sensitivity." We will refer to this value as "sensitivity" below.

Subsequently, the Allan deviation was calculated for the measured 300 spectra at temperatures of 55 °C and 276 °C, taking its value at a time of 7 seconds $\partial T_{Allan}(7\text{s})$. The procedure was repeated for different $\lambda_{\text{inflection}}, W$ with steps of 0.1 nm and 1 nm, respectively (Figure 3). The optimal filter values are obtained at the ZPL slopes (ZPL centers at 602.8 nm and 608.0 nm, respectively). Operating on the left slope (λ < ZPL) is not very practical, as it requires a very narrow filter with sharp boundaries. It is more convenient to work on the right slope, and, as seen, the values for it are $\partial T^{276}_{Allan}(7s) = 0.85\ ^{\circ}\mathrm{C}$ at $\lambda_{\text{inflection}} / W = 612.5/20$ nm and $\partial T^{55}_{Allan}(7s) = 0.6\ ^{\circ}\mathrm{C}$ at $\lambda_{\text{inflection}} / W = 607.44/5$ nm. The obtained improvement is comparable to the approximation method proposed in [5]. Optimizations were performed for the four window boundaries for $R^4_{ideal}(T) = \sum_{602.2\le\lambda\le612.4} I(\lambda) \Big/ \sum_{607.2\le\lambda<636.1} I(\lambda)$ with the sensitivity for both temperatures $\partial T^{276\&55}_{Allan}(7s) < 0.8\ ^{\circ}\mathrm{C}$ (and $\Delta T_{mean} \sim 0.7\ ^{\circ}\mathrm{C}$ is comparable to the stabilization and temperature setting error in our experimental setup).

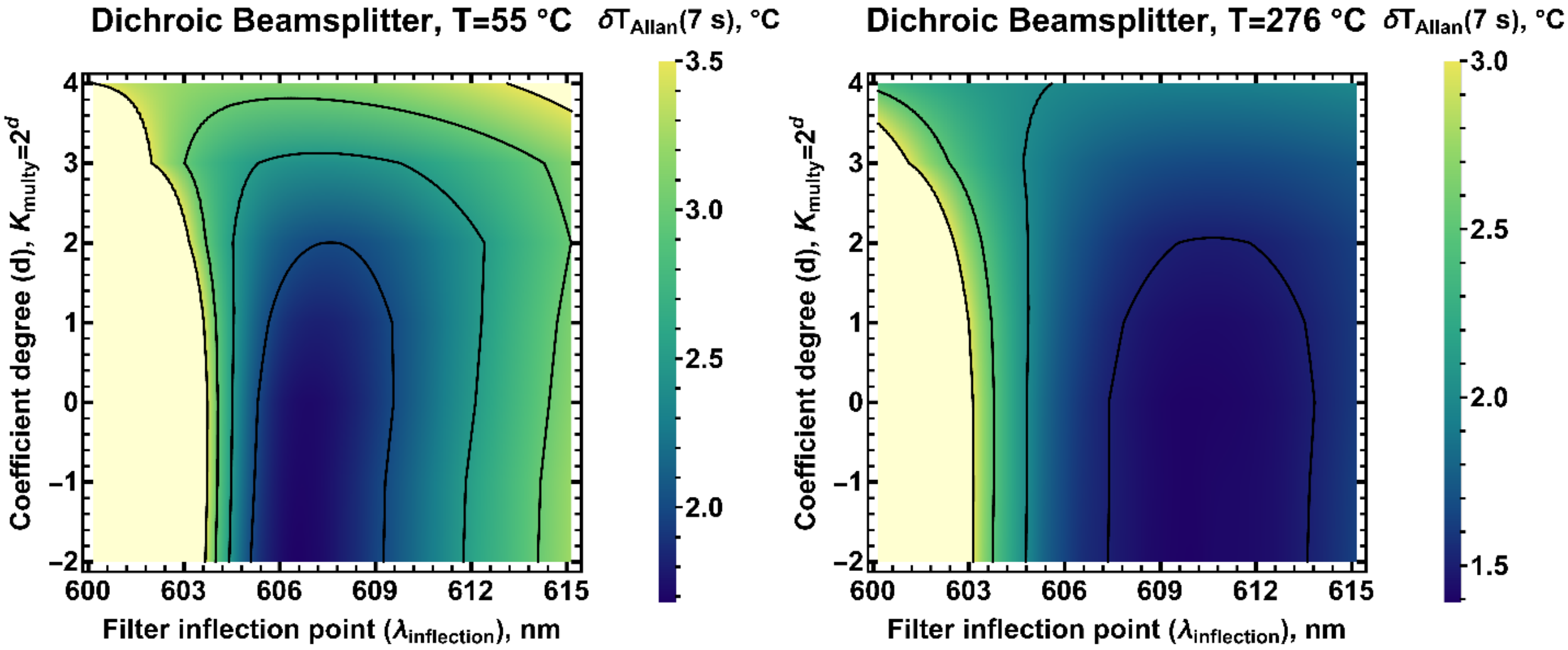

*Figure 4 Allan deviation of a temperature taken at 7 s ($\partial T_{Allan}(7\mathrm{s})$) as a function of the parameters calculated using data from a real dichroic mirror.*

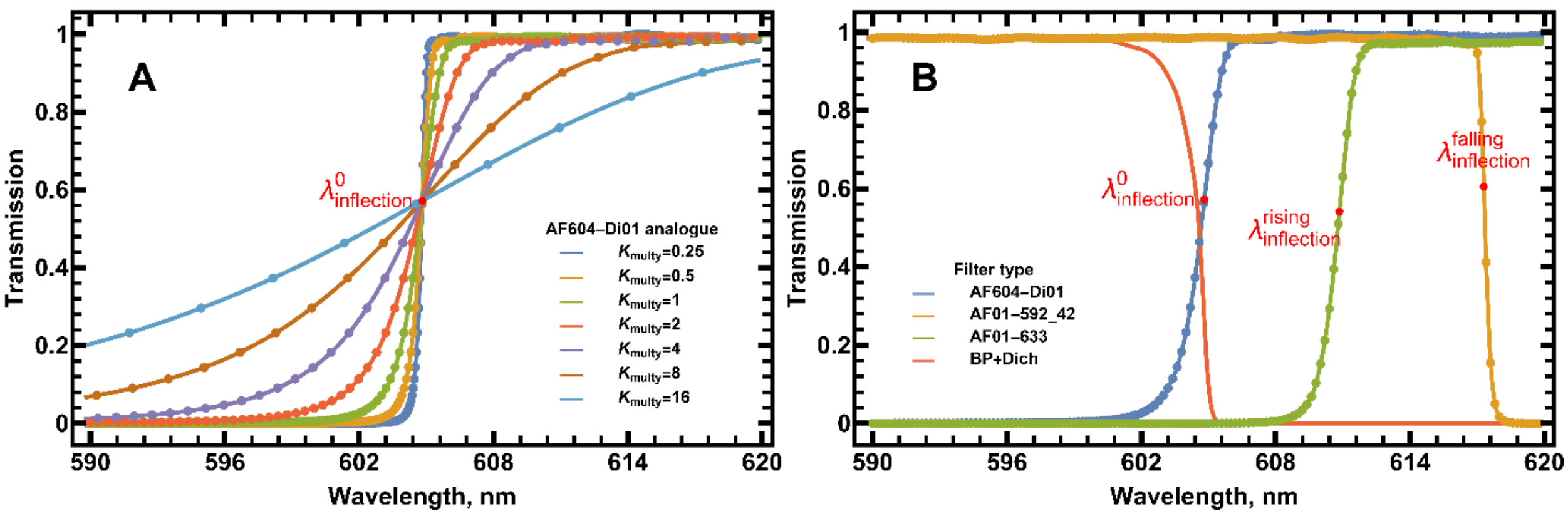

*Figure 5 A) Transmission functions of AF604-Di01 filter analogues with a multiplication factor $K_{multy} = 2^d$. B) Transmission and inflection points for all the studied filters.*

It is practically advantageous if the filter simply splits the spectrum without restricting it to a specific window. Therefore, a graph was constructed in which the sum was taken over the entire available spectrum $\lambda \in 577-720\,\text{nm}$. This corresponds to a setup with a single dichroic mirror that simply divides the spectrum. As can be seen, the obtained values are 2 (3) times higher than those obtained at W = 20(5) nm. However, sensitivity remains better at 276 °C compared to the approximation of the ZPL width obtained from the same data (Figure 6). The primary advantage of the current method lies in its compactness and ease of implementation.

It is also worthwhile to consider how sensitivity changes when using a real filter. For this purpose, the dichroic beam splitter Semrock AF604-Di01 was selected as an example. For simplicity, it was assumed that the filter did not absorb, and consequently, the signal reached one of the photodiodes. The transmission characteristics provided by the manufacturer were interpolated as a function of wavelength $Tr(\lambda)$, allowing us to determine the inflection point by finding the zero of the second derivative $\lambda^0_{\text{inflection}} = 604.815$ nm. Additionally, to assess the impact of the filter's transition sharpness, the $K_{multy} : Tr_{multy}(\lambda) = Tr\left(\dfrac{\lambda - \lambda^0_{\text{inflection}}}{K_{multy}} + \lambda^0_{\text{inflection}}\right)$ multiplication factor was introduced, which varied as a power of two $K_{multy} = 2^d$ (Figure 5), if $\lambda$ exceeded the interpolation limits, the transmission was considered to be 1 and 0, respectively. The computational experiment described above was repeated with the signal ratios on the photodiodes, varying $\lambda_{\text{inflection}}$:

$$R_{real}(T) = \sum_{577 \le \lambda \le 720} \frac{I(\lambda) Tr_{multy}\left(\lambda - \lambda_{\text{inflection}} + \lambda^0_{\text{inflection}}\right)}{I(\lambda)\left[1 - Tr_{multy}\left(\lambda - \lambda_{\text{inflection}} + \lambda^0_{\text{inflection}}\right)\right]}, \tag{2}$$

the results of which are presented in Figure 4. At $d = 0$ (the analogs of a real filter AF604-Di01), the sensitivity at $T = 55\ °C$ and 7 second integration time decreased by half $\partial T_{Allan}^{55}(7s) = 1.7\ ℃$ than when operating the sensor to determine the ZPL width, but it remained twice as effective at $T = 276\ °C$: $\partial T_{Allan}^{276}(7s) = 1.4\ ℃$. It is important to note, that as temperature changes, optical inflection point changes as well. This change is nevertheless is relatively small. For example, if scenario 1 is used to optimizes sensitivity at 55°C filters AF604-Di01 (d=0) inflection point need to be set to 606.94 nm. With the same inflection point the sensitivity at temperature 300 °C will be only 10% different from optimal. The deterioration in slope by a factor of 16 corresponds to a twofold decrease in sensitivity. However, designing a filter with a sharper transition $(d = -1, -2)$ is not particularly beneficial, as other factors limit the sensitivity relative to the ideal filter. Besides the slope, the ideal filter differs in that the transmission and reflection do not equal 100%, which explains the observed sensitivity limitation when $d \le 0$.

A slight improvement in sensitivity might be achieved by replacing the long-pass filter with a band-pass filter, shifted as shown in (Figure 2-2). The band-pass filter Semrock AF01-592/42-25 was tested as an example. Following a similar procedure, the spectral position of the inflection point of transmission of the band-pass filter $\lambda_{\text{inflection}}^{\text{falling}} : Tr''\left(\lambda_{\text{inflection}}^{\text{falling}}\right) = 0, Tr'\left(\lambda_{\text{inflection}}^{\text{falling}}\right) < 0$ was varied from 600.5 to 621.5 nm, along with the inflection point of the dichroic mirror at $d = 0$. This setup allows for a 20% improvement in sensitivity and proved to be insensitive to the spectral position $621.5 \ge \lambda_{\text{inflection}}^{\text{falling}} \ge 608.5\ nm$ (see Supporting Information). The optimal position of the dichroic mirror is close to that obtained in Figure 4. A comparison of the sensitivity achieved by different measurement methods is presented in Figure 6.

| Temperature range | Measurement scheme | 1 | 2 | 3 |
|---|---|---|---|---|
| $T \in [195, 285]\ ℃$ | $\max_T \lvert p(3, R(T)) - p(1, R(T)) \rvert$ | 5.2 ℃ | 4.2 ℃ | 2.7 ℃ |
| $T \in [25, 115]\ ℃$ | $\max_T \lvert p(3, R(T)) - p(1, R(T)) \rvert$ | 1 ℃ | 1.15 ℃ | 2.1 ℃ |
| $T \in [195, 285]\ ℃$ | $\max_T \lvert p(3, R(T)) - p(2, R(T)) \rvert$ | 0.85 ℃ | 0.55 ℃ | 0.12 ℃ |
| $T \in [25, 115]\ ℃$ | $\max_T \lvert p(3, R(T)) - p(2, R(T)) \rvert$ | 0.05 ℃ | 0.04 ℃ | 0.13 ℃ |

*Table A The estimated temperature error associated with non-linearity for different temperature ranges (as the maximum difference between the approximation of a ratio signal R(T) by polynomials of different degrees p(degree, R(T))) at the maximum sensitivity for the specified measurement scheme option: 1 – AF604-Di01, 2 – AF604-Di01 and AF01-592/42-25, 3 – AF01-592/42-25 and AF01-633/40-25.*

An alternative approach could involve using two band-pass filters (or a band-pass and a long-pass filter), as shown in the third option in Figure 2-3. This scheme has the advantage of not requiring precise angle alignment of the dichroic mirror (splitter). The correct band position can be set during the filter manufacturing process, which is typically simpler and more cost-effective for a band-pass filter than for a dichroic mirror. For an example, the band-pass filters Semrock AF01-592/42-25 and AF01-633/40-25 were selected, for which falling $\lambda_{\text{inflection}}^{\text{falling}}$ and rising $\lambda_{\text{inflection}}^{\text{rising}} : Tr''\left(\lambda_{\text{inflection}}^{\text{rising}}\right) = 0, Tr'\left(\lambda_{\text{inflection}}^{\text{rising}}\right) > 0$ points of inflection were varied (see Supporting Information). The obtained sensitivities at $\lambda_{\text{inflection}}^{\text{rising}} = 609.5\,(604.4)$ nm and $\lambda_{\text{inflection}}^{\text{falling}} = 612.3\,(621.6)$ nm is $\partial T_{Allan}^{276}(7s) = 1.5$ ℃ and $\partial T_{Allan}^{55}(7s) = 0.85$℃ (were multiplied by $\sqrt{2}$, as the beam splitter will divide the signal) (Figure 6).

Another critical parameter for the temperature sensor is linearity. Therefore, the obtained data $R_{real}(T)$ (see Figure 7) were also approximated using second-degree $p\left(2, R(T)\right)$ and first-degree polynomials $p(1, R(T))$. The maximum deviation $\max_T \left| p\left(3, R(T)\right) - p\left(1, R(T)\right) \right| : T \in [25, 370]$℃ can be quite significant – about ten degrees. Nevertheless, it is useful to separately examine the temperature regions $T^F \in [25,115]$ ℃ and $T^S \in [195, 285]$ ℃ (see Supporting Information) where the greatest deviation between linear and cubic interpolation occurs at optimal sensitivity parameters (for each scheme presented in Figure 2). Table A indicates that the second scheme option exhibits lower non-linearity. For higher temperatures, however, non-linear calibration is recommended. Consequently, a comparison between second-degree and third-degree polynomial interpolations was performed, demonstrating that calibration accounting for quadratic non-linearity is adequate for many applications.

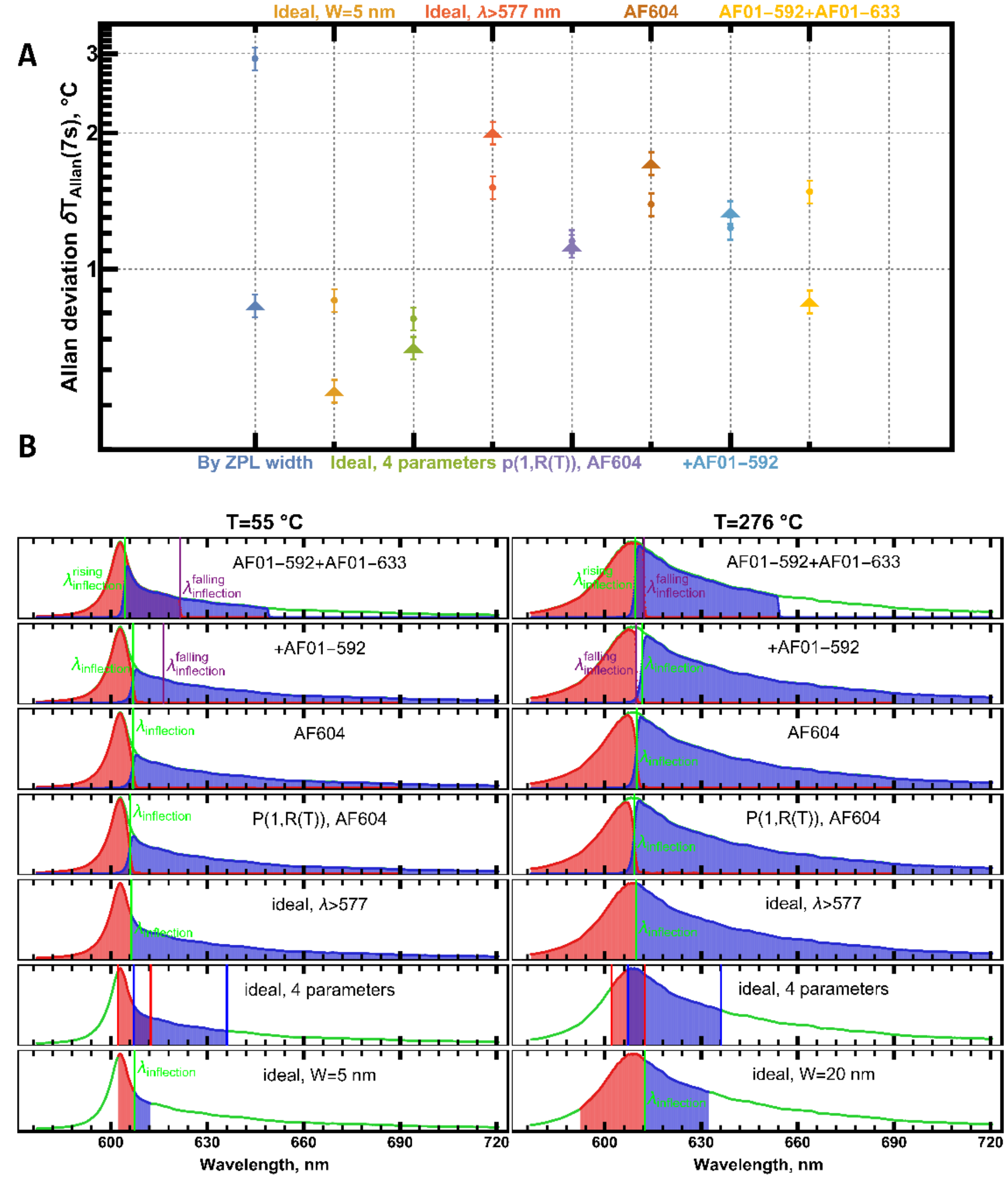

*Figure 6 A) Comparison of sensitivity for different measurement methods and filter parameters (triangle – 55 °C, circle – 276 °C). Error bar corresponds to statistical error of Allan deviation estimation. B) The mean spectra (green) and their parts that were used to calculate the signal ratios (total of red/ total of blue).*

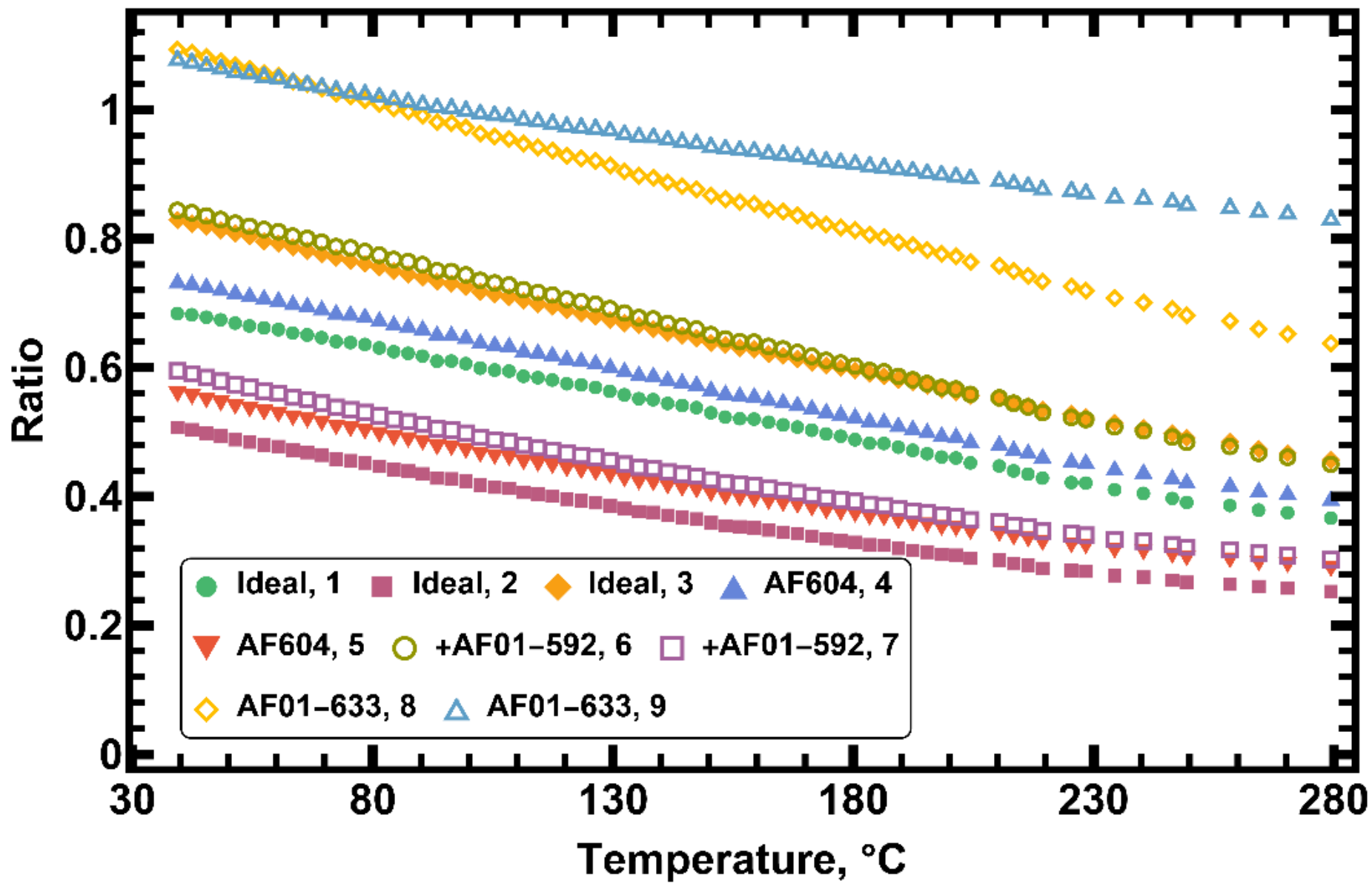

*Figure 7 Example of the relationship between signal ratios* $R(T)$ *and temperature for optimal sensitivity. In the legend, the following parameters of the filters were used: 1: an ideal dichroic mirror with the inflection point of filter transmission* $\lambda_{\text{inflection}} = 609.7\text{ nm}$*; 2: an ideal dichroic mirror,* $\lambda_{\text{inflection}} = 606.4\text{ nm}$*; 3: an ideal filter with 4 parameters (see text) which define bands in each arm; 4: an analog of a real filter AF604 with the inflection point of filter transmission* $\lambda_{\text{inflection}} = 609.9\text{ nm}$*; 5: the same filter, but with* $\lambda_{\text{inflection}} = 606.9\text{ nm}$*; 6: an analog of a real filter AF604 with* $\lambda_{\text{inflection}} = 611.6\text{ nm}$ *and an analog of A01-592 with a falling point of inflection with* $\lambda_{\text{inflection}}^{\text{falling}} = 609.8\, nm$*; 7: the same but with* $\lambda_{\text{inflection}} / \lambda_{\text{inflection}}^{\text{falling}} = 606.9 / 616.5\, nm$*; 8: an analog of two band-pass filters with the transmission rising/falling point of inflection* $\lambda_{\text{inflection}}^{\text{rising}} / \lambda_{\text{inflection}}^{\text{falling}} = 609.4 / 612.3\, nm$*; 9: the same with* $\lambda_{\text{inflection}}^{\text{rising}} / \lambda_{\text{inflection}}^{\text{falling}} = 604.4 / 612.3\, nm$*.*

## IV. CONCLUSION

This study examined a simplified design for an optical temperature sensor based on GeV centers, enabling the replacement of a spectrometer with a dichroic mirror, optionally combined with an additional band-pass filter. Temperature was determined by measuring the signal ratio between two photodiodes positioned at the transmission and reflection beams after the dichroic mirror. An alternative configuration using two band-pass filters and a beam splitter was also proposed. The expected sensitivities with optimally selected filters were comparable to, and in some cases exceeded, those achieved with the traditional method of measuring the ZPL width, particularly at elevated temperatures. Various commercially available filters and simplified configurations were evaluated in terms of potential sensor sensitivity. The study demonstrated the sensitivity and nonlinearity dependence of the temperature sensor on the position of the filter band. For temperature measurements near room temperature, the traditional ZPL width analysis slightly outperformed the dichroic mirror or band-pass filter method. However, at temperatures around 300 °C, the dichroic mirror-based method surpassed the traditional approach, showing the estimated sensitivity improvement of approximately a factor of 2, even with relatively moderate dichroic or band-pass filters. This is due to the ZPL width-based method's performance declining by about a factor of 3 at higher temperatures, while the performance of the dichroic mirror and band-pass filter methods remains stable. Furthermore, the latter approach significantly reduces sensor costs and enables a more compact design, making it an attractive option for industrial applications.

## V. ACKNOWLEDGEMENTS

This research was supported by a grant of the Ministry of Science and Higher Education of the Russian Federation No. 075-15-2024-556.